 \documentclass[pmlr,twocolumn,10pt]{jmlr} % W&CP article

\usepackage{booktabs}
\usepackage{siunitx}
\newcommand{\equal}[1]{{\hypersetup{linkcolor=black}\thanks{#1}}}

\theorembodyfont{\upshape}
\theoremheaderfont{\scshape}
\theorempostheader{:}
\theoremsep{\newline}

\jmlrvolume{LEAVE UNSET}
\jmlryear{2023}
\jmlrsubmitted{LEAVE UNSET}
\jmlrpublished{LEAVE UNSET}
\jmlrworkshop{Machine Learning for Health (ML4H) 2023} % W&CP title

\title[Osteoporosis Prediction from Hand and Wrist X-rays using Image Segmentation and Self-Supervised Learning]{Osteoporosis Prediction from Hand and Wrist X-rays using Image Segmentation and Self-Supervised Learning}

\author{%
  \Name{} \Email{}\\
  \addr
 }
 
\author{%
  \Name{Hyungeun Lee}\equal{These authors contributed equally}
  \Email{didls1228@hanyang.ac.kr}\\
  \Name{Ung Hwang}\footnotemark[1]
  \Email{gogowinner@hanyang.ac.kr}\\
  \Name{Seungwon Yu}\footnotemark[1]
  \Email{ahddl1324@hanyang.ac.kr}\\
  \addr Department of Electronic Engineering, Hanyang University, Seoul, Korea
  \AND
  \Name{Chang-Hun Lee}
  \Email{drlch79@hanyang.ac.kr}\\
  \addr Department of Orthopedic Surgery, Hanyang University, Seoul, Korea
  \AND
  \Name{Kijung Yoon}\equal{Corresponding author}
  \Email{kiyoon@hanyang.ac.kr}\\
  \addr Department of Electronic Engineering, Hanyang University, Seoul, Korea\\
  \addr Department of Artificial Intelligence, Hanyang University, Seoul, Korea
 }

\begin{document}

\maketitle

\begin{abstract}
Osteoporosis is a widespread and chronic metabolic bone disease that often remains undiagnosed and untreated due to limited access to bone mineral density (BMD) tests like Dual-energy X-ray absorptiometry (DXA). In response to this challenge, current advancements are pivoting towards detecting osteoporosis by examining alternative indicators from peripheral bone areas, with the goal of increasing screening rates without added expenses or time. In this paper, we present a method to predict osteoporosis using hand and wrist X-ray images, which are both widely accessible and affordable, though their link to DXA-based data is not thoroughly explored. Initially, our method segments the ulnar, radius, and metacarpal bones using a foundational model for image segmentation. Then, we use a self-supervised learning approach to extract meaningful representations without the need for explicit labels, and move on to classify osteoporosis in a supervised manner. Our method is evaluated on a dataset with 192 individuals, cross-referencing their verified osteoporosis conditions against the standard DXA test. With a notable classification score (AUC=0.83), our model represents a pioneering effort in leveraging vision-based techniques for osteoporosis identification from the peripheral skeleton sites.

\end{abstract}

\section{Introduction}
\label{sec:intro}
Osteoporosis is a common bone ailment characterized by reduced bone mineral density (BMD) or bone mass loss, leading to bones becoming fracture-prone and structurally compromised. Given its prevalence and far-reaching impact, there is a pressing need for preemptive risk assessment, early diagnosis, and effective preventive actions. While computed tomography (CT) and magnetic resonance imaging (MRI) have demonstrated potential in BMD estimation and osteoporosis screening \citep{chou2017vertebral, pickhardt2013opportunistic, gausden2017opportunistic}, their clinical use is limited due to concerns about radiation exposure and associated costs. At present, the dual energy X-ray absorptiometry (DXA) is recognized as a standard and reliable instrument for osteoporosis detection and BMD analysis. Nonetheless, DXA comes with its own set of challenges: the proficiency of the operator can influence the results, the patient's posture during the test can skew accuracy, and in obese patients with significant fat mass, BMD might be overestimated \citep{watts2004fundamentals,messina2015prevalence}. 

As a solution, recent studies have highlighted the potential of X-rays, often performed during routine medical visits, as a valuable tool for collecting more comprehensive data. This approach can be particularly useful for osteoporosis detection \citep{yamamoto2020deep,hsieh2021automated}, leveraging the widespread availability of X-ray data from patients who have not been specifically screened for the condition. This method eliminates the need for expensive DXA equipment, making it a more cost-effective alternative. X-rays are also more patient-friendly due to their significantly lower radiation levels compared to DXA. Among the various techniques, the 2nd metacarpal cortical index (2MCI) from hand X-ray images has gained attention as a promising biomarker for osteoporosis screening \citep{schreiber2017simple, patel20202nd}. The metacarpal bone's unique features, such as cortical thickness and the porosity of the cancellous bone tissue, are key in identifying osteoporosis. Nonetheless, this approach is not without its challenges, as it requires manual measurement of the metacarpal bone dimensions, a process that can be time-consuming.

\begin{figure*}[t]
 % Caption and label go in the first argument and the figure contents
 % go in the second argument
\floatconts
  {fig:sam}
  {\caption{\textbf{Illustration of the proposed osteoporosis prediction framework.} \textbf{(a)} Raw X-ray images marked with point prompts (colored circles). These prompts guide SAM in generating output masks (colored regions), leading to the derivation of specific bone segments (colored squares). \textbf{(b)} For each of the seven bone segments, we extract 2 global views and 4 local views, courtesy of our refined multi-crop augmentation approach. \textbf{(c)} These augmented inputs are directed towards pretext training through a self-supervised learning system. \textbf{(d)} The backbone classification encoder network (orange trapezoid) is then repurposed to tackle the downstream classification task. \textbf{(e)} The same global and local views from the original multi-crop augmentation method \citep{caron2020unsupervised}.}}
  {\includegraphics[width=\linewidth]{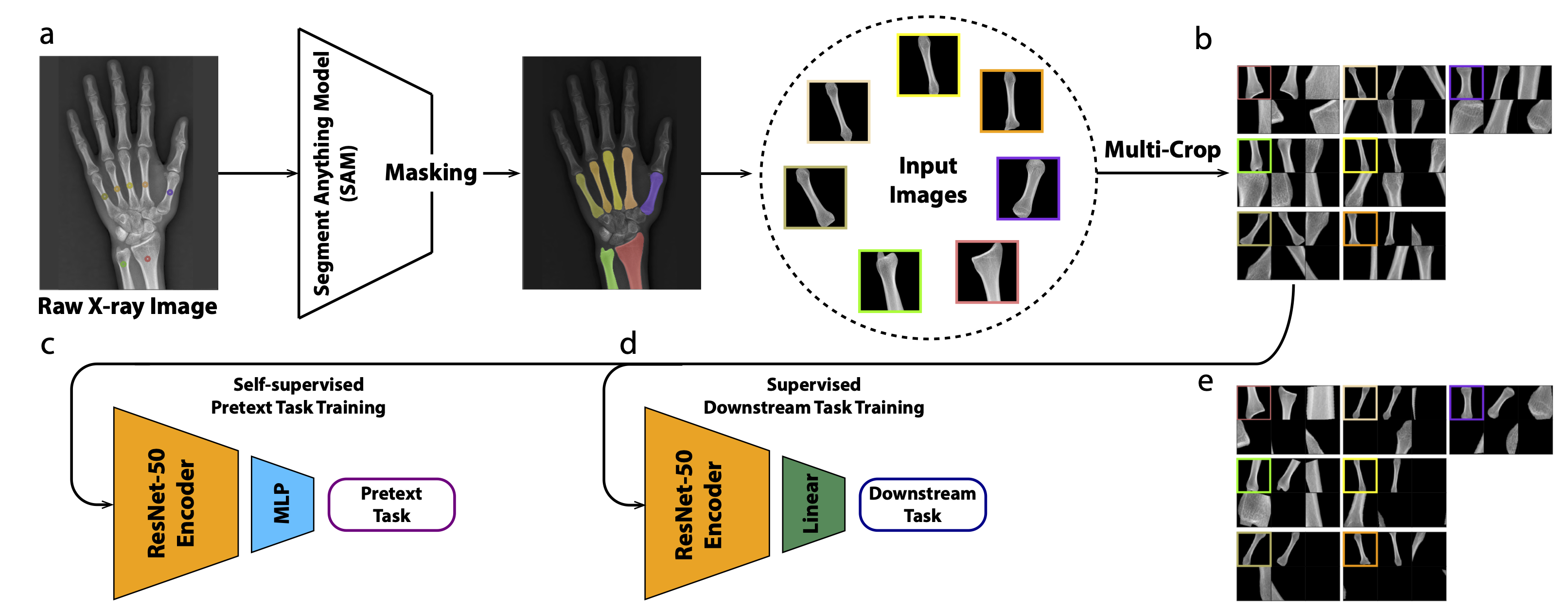}}
\end{figure*}

Recent strides in deep learning techniques, showcasing diagnostic precision comparable to human experts in numerous medical fields \citep{huang2023self}, motivate us to tackle the osteoporosis detection challenge taking advantage of existing research. We do this by utilizing a deep learning approach on hand and wrist X-ray images, eliminating the necessity for manual 2MCI calculations. In this study, our first step is to curate the training data by isolating significant bones in the hand and wrist, especially the metacarpals, relevant for osteoporosis screening. We achieve this bone separation using a foundational image segmentation tool known as the Segment Anything Model (SAM) \citep{kirillov2023segany}. Following that, we adopt a self-supervised learning (SSL) framework for the preliminary training of our models. This method allows the models to learn meaningful representations without relying on specific labeled datasets. Such an SSL strategy enhances the model's generalization capabilities, even in scenarios of limited datasets, as is the case in our study with $N=192$ participants. Moreover, we expand upon the multi-crop augmentation strategy \citep{caron2020unsupervised}, tailored for hand X-ray images, to improve the efficiency of our self-supervised models.

In summary, our contributions encompass three key aspects:
\begin{itemize}
    \item We employ a deep learning approach leveraging peripheral radiographic images over traditional central DXA measurements for osteoporosis prediction, obviating the need for 2MCI metric. This novel strategy could pave the way for more convenient osteoporosis screenings.
    \item We propose an SSL approach centered on segments of hand and wrist bones, enabling our models to distill critical features prior to the subsequent osteoporosis prediction phase. This method demonstrates enhanced outcomes when compared with the conventional supervised models.
    \item Our study suggests that the proposed method achieves clinically valuable osteoporosis screening results even under the low data regime. We anticipate potential improvements by integrating more unlabeled radiographic data in future iterations.
\end{itemize}

\begin{table}[t]
\floatconts
  {tab:dataStat}
  {\caption{We split the 1636 bone segments into training, validation, and testing sets following an 8:1:1 ratio. The validation set assists in monitoring the model's downstream task performance, while the test set evaluates the quality of osteoporosis predictions. In each split, approximately 29\% of the subjects have osteoporosis consistently across the datasets.}}
  {%
    \begin{tabular}{@{}llll@{}}
\toprule
\# samples & Train & Val & Test \\ \midrule
Osteoporosis       & 372   & 46  & 48   \\
Normal & 935   & 116 & 119  \\
Total        & 1307  & 162 & 167  \\ \bottomrule
\end{tabular}%
  }
\end{table}

\section{Methods}
\label{sec:methods}
In this section, we first introduce the way of generating segmentation masks for targeted bones (Sec. \ref{sec:mask}), multi-crop data augmentation methods (Sec. \ref{sec:aug}), present strategies for self-supervised learning on the segmented image inputs (Sec. \ref{sec:ssl}), and then detail the experimental procedure (Sec. \ref{sec:exp_details}). 

\subsection{Segmentation Mask for Target Bones}
\label{sec:mask}
Our initial assumption posits that raw X-ray images may not be the optimal input for our backbone residual neural networks. This is because these raw radiographs, averaging 2515(W) by 3588(H) in size, are typically too large for most conventional vision models. Merely resizing them might lead to a loss of critical features like bone texture and microarchitecture \citep{lespessailles2008clinical, zheng2016bone,sebro2022machine} which are essential for predicting osteoporosis. Furthermore, drawing from prior studies that highlight a notable correlation between 2MCI and BMD measurements \citep{schreiber2017simple, patel20202nd}, we extract smaller individual bones from the raw X-rays of hands and wrists.

To facilitate this, we employ SAM \citep{kirillov2023segany} as a foundation model for image segmentation to generate specific bone masks, as illustrated in Figure \ref{fig:sam}a. SAM allows us to more precisely separate each target bone by setting positive point prompts within the desired region and adding negative point prompts outside that area. The resultant output masks are then used to produce targeted bone image segments using a simple Hadamard multiplication (Figure \ref{fig:sam}a, right). In the end, SAM is applied to segment 192 X-ray images, yielding a total of 1636 images for model inputs, comprising the ulna, radius, and five metacarpals (more details in Table \ref{tab:dataStat}). It is worth mentioning that there were occasional deviations resulting in fewer than seven target image samples. This was due to irregular hand positioning or the presence of accessories or equipment on the patient's hand.

\subsection{Multi-Crop Data Augmentation}
\label{sec:aug}
The bone segments derived in the previous step exhibit variations in rotational angles (Figure \ref{fig:sam}a, right), primarily due to the inherent anatomical structure of the hand and wrist and the hand position. To ensure our model can robustly handle such variations, we apply multiple data augmentation techniques, including random rotations between -20 to 20 degrees and random horizontal/vertical flips. Additionally, we incorporate the multi-crop data augmentation method \citep{caron2020unsupervised} to benefit from efficient pretext task training in Section \ref{sec:ssl}. This method samples images of two distinct resolutions: randomly seletected two global views ($224\times 224$) and four local views ($96\times 96$).

However, conventional multi-crop augmentation, often applied to standard image classification benchmarks or natural images, ensures each cropped segment remains feature-rich. In contrast, our specific bone segments frequently have vast areas of zero pixels, a result of the segmentation process. This characteristic persists even in original radiographs before applying SAM; for a visual reference, see Figure \ref{fig:sam}e. Consequently, some samples might entirely lack bone content.

To address this issue, we adapt the multi-crop augmentation technique. Our enhanced approach ensures that both global and local view outputs consistently possess a minimum threshold of non-zero pixels. We achieve this consistency using iterative rejection sampling until our criteria are satisfied. This guarantees the utility and relevance of each cropped image in subsequent phases (Figure \ref{fig:sam}b). For ensuring meaningful content in the cropped segments, we establish a 10\% non-zero pixel threshold, minimizing the risk of feature loss.

\subsection{Self-Supervised Learning}
\label{sec:ssl}
In this study, we evaluate four prominent SSL methods: SimCLR \citep{chen2020simple}, SupCon \citep{khosla2020supervised}, SwAV \citep{caron2020unsupervised}, and VICReg \citep{bardes2021vicreg} all fall under the umbrella of self-supervised learning, emphasizing the extraction of meaningful patterns from data without using explicit labels. A cornerstone for all these methods is the use of data augmentation strategies, ensuring that the neural networks learn robust and generalizable features. By designing tasks that force the models to recognize or distinguish between differently augmented views of the same data, these methods effectively train models to focus on inherent data structures. Moreover, all four techniques share the overarching goal of maximizing consistency between these augmented views in a shared embedding space, ensuring that semantically similar data points (or their augmented versions) are mapped closer together.

SimCLR \citep{chen2020simple} focuses on contrasting different augmented versions of the same instance against other instances, while SupCon \citep{khosla2020supervised} enhances this paradigm by also bringing instances of the same class closer in the embedding space. SwAV \citep{caron2020unsupervised} introduces a \textit{swapped prediction mechanism} avoiding the need for negative pair comparisons, making it computationally more efficient. In contrast, VICReg \citep{bardes2021vicreg} stands out by imposing variance, invariance, and covariance regularization on the latent space, ensuring the captured features are diverse and not merely based on trivial patterns. Each method, thus, introduces its unique strategy to address the challenges of self-supervised learning in its way. The exploration of effective self-supervised learning methods in peripheral radiographic images remains nascent, making our experimental results useful for clinical tasks in other medical domains.

\subsection{Experimental Details}
\label{sec:exp_details}
Our experiments involve two distinct training stages. In the initial stage, known as the pretext task, we train the ResNet-50 encoder on augmented views that are generated automatically. This step enables the encoder to extract meaningful features from the images without needing labels provided by humans. Following this, in the downstream task, the pre-trained encoder undergoes fine-tuning on a dataset that is labeled specifically for our target task of classification. In this phase, the encoder serves as an efficient feature extractor, and we train only its final linear layer for accurate osteoporosis prediction. For the optimization process, we employ the LARC optimizer \citep{you2017large}, along with a cosine annealing scheduler \citep{loshchilov2016sgdr} to adjust the learning rates. The entire training process is executed with a batch size of 128.

For accurate target labels, we cross-reference our data with DXA scans. DXA operates by assessing the variation in the attenuation (absorption) of X-rays by the bones to determine the bone mineral density. BMD quantifies the concentration of calcium and other minerals in a specific bone segment. The obtained BMD value is instrumental in computing the $T$-score, a metric that benchmarks a patient’s BMD against the average BMD of a healthy 30-year-old adult. A $T$-score in the negative spectrum denotes a BMD that is below the average, while a positive $T$-score signifies an above-average BMD. In our study, we utilize the $T$-score as a criterion to annotate the X-ray images. An X-ray image is tagged as indicative of osteoporosis if the associated $T$-score falls below $-2.5$; otherwise, it is classified as normal.

\begin{table*}[t]
\floatconts
  {tab:multicropresult}
  {\caption{Results of osteoporosis prediction on radiographic test images using a standard supervised learning approach compared with four SSL techniques, in conjunction with two multi-crop augmentation strategies. Ext. multi-crop represents the extended strategy of ensuring the 10\% non-zero pixel threshold. All numbers presented are the average of 3 random trials.}}%
{
\resizebox{\textwidth}{!}{
\begin{tabular}{c|ccc|ccc|ccc}
\hline
       & \multicolumn{2}{c}{AUC}                                         & $\Delta$              & \multicolumn{2}{c}{F1-score}                                        & $\Delta$              & \multicolumn{2}{c}{Accuracy}                                  & $\Delta$              \\ \hline
SSL($\downarrow$)/ \text{Augmentation}($\rightarrow$) & multi-crop \citeyearpar{caron2020unsupervised}    & ext. multi-crop     & \multicolumn{1}{l|}{}  & multi-crop \citeyearpar{caron2020unsupervised}    & ext. multi-crop     & \multicolumn{1}{l|}{}  & multi-crop \citeyearpar{caron2020unsupervised}    & ext. multi-crop     & \multicolumn{1}{l}{} \\ \hline
Supervised & 0.524 $\pm$ 0.011 & 0.527 $\pm$ 0.004  & +0.003& 0.447 $\pm$ 0.000 & 0.447 $\pm$ 0.000 & +0.000& 28.74 $\pm$ 0.00 & 28.74 $\pm$ 0.00 & +0.00\\
\hline
SimCLR \citeyearpar{chen2020simple} & 0.809 $\pm$ 0.011 & \textbf{0.827} $\pm$ 0.003 & +0.016               & 0.500 $\pm$ 0.014 & \textbf{0.655} $\pm$ 0.034 & +0.155               & 79.24 $\pm$ 0.69 & \textbf{79.84} $\pm$ 0.35 & +0.60                 \\
SupCon \citeyearpar{khosla2020supervised} & 0.787 $\pm$ 0.013 & 0.813 $\pm$ 0.005 & +0.026               & 0.562 $\pm$ 0.040 & 0.611 $\pm$ 0.025 & +0.049               & 77.45 $\pm$ 0.35 & 74.85 $\pm$ 5.12 & -2.60                 \\
SwAV \citeyearpar{caron2020unsupervised}   & 0.778 $\pm$ 0.023 & 0.736 $\pm$ 0.062 & -0.041               & 0.400 $\pm$ 0.096 & 0.567 $\pm$ 0.058 & +0.167               & 76.45 $\pm$ 0.69 & 67.07 $\pm$ 6.59 & -9.38                 \\
VICReg \citeyearpar{bardes2021vicreg} & 0.806 $\pm$ 0.015 & 0.812 $\pm$ 0.018 & +0.006               & 0.522 $\pm$ 0.084 & 0.634 $\pm$ 0.031 & +0.112               & 77.05 $\pm$ 0.69 & 74.45 $\pm$ 3.07 & -2.60                 \\ \hline
\end{tabular}
}
}
\end{table*}

\section{Results}
Table \ref{tab:multicropresult} displays the osteoporosis prediction results, measured through AUC, F1-score, and accuracy, for both supervised and self-supervised learning approaches. The first striking observation is the supervised model's dismal accuracy at 28.74\%.  With its average accuracy closely mirroring the percentage of osteoporotic patients and a near-zero validation F1-score (Figure \ref{fig:supervised}), there is an implication that the supervised model may consistently lean toward a single decision regardless of the radiographic images fed into it. This result highlights the supervised approach's marked inability to learn meaningful representations for distinguishing osteoporosis from normal cases. We have thoroughly investigated this issue by employing various tactics, including tuning hyperparameters, applying regularization techniques, and experimenting with data augmentation methods. Despite these efforts, the problem remains unresolved. We attribute this challenge to the intricate interplay between the limited sample size and the complexity of the backbone network, which appears to adversely impact the model’s performance in a supervised context.

On the other hand, all SSL techniques demonstrate decent efficacy in discerning between positive and negative instances, albeit with potential for enhancement. Remarkably, of the four SSL methods, SimCLR in tandem with our refined multi-crop augmentation emerges superior to other baselines, despite not being the latest algorithm. Moreover, our tailored multi-crop technique, applied to radiographic images, yields promising outcomes, underscoring the significance of pinpointing the pertinent bone regions. 

In essence, the SSL-centric osteoporosis classification model reports an AUC of 0.827, an F1-score of 0.655, and an accuracy rate of 79.84\%. These metrics mark a pivotal leap in osteoporosis imaging classification, affirming the viability of using hand and wrist X-rays for preliminary osteoporosis detection.

\section{Conclusion}
% \todo{Summarize the main contributions mentioned from bullet point list  in section 1}
In this study, we explored the potential of using a self-supervised learning technique with hand and wrist radiographic images to determine osteoporosis status. Utilizing the robust point prompt capabilities of SAM, we effectively produced segmented bone images and implemented our unique extended multi-crop image augmentation method to enhance the model's predictive accuracy. While this approach is in its early stages, we are actively engaging with tens of thousands of unlabeled radiographs. We believe it holds promise for leading to more precise and user-friendly osteoporosis screenings, which we aim to further develop in the future.

\acks{This work was supported in part by the Institute of Information \& communications Technology Planning \& Evaluation (IITP) grant (No.2020-0-01373, Artificial Intelligence Graduate School Program (Hanyang University)) funded by the Ministry of Science and ICT (MSIT), and in part by the National Research Foundation of Korea (NRF) under Grant NRF-2020R1A2C2101353}

\newpage
\bibliography{jmlr-sample}

\newpage
\appendix

\section{Multi-crop data augmentation strategy}
The multi-crop data augmentation strategy, as proposed in \citet{caron2020unsupervised} involves utilizing two standard resolution global crops along with $V$ additional low-resolution crops that capture local regions of the image. The authors demonstrate that this augmentation strategy enhances the performance of various self-supervised learning models while maintaining memory efficiency.
In all our pretext tasks, We employ the generalized loss term proposed in \citet{caron2020unsupervised} for both strategies as follows: 
\begin{displaymath}
L(\mathbf{z}_{t_1}, \mathbf{z}_{t_2}, \ldots, \mathbf{z}_{t_{V+2}}) = \sum_{i \in \{1, 2\}}{\sum_{v=1}^{V+2}{\mathbf{1}_{v \neq i}l(\mathbf{z}_{t_v}, \mathbf{z}_{t_i})}}
\end{displaymath}
In SwAV case, we regard $\mathbf{z}_{t_i}$ as $\mathbf{q}_{t_i}$, which is a code computed from the iterative Sinkhorn-Knopp algorithm \citep{cuturi2013sinkhorn}. The formula represents a loss function used in our pretext tasks. Here's a breakdown of its components:
\begin{itemize}
    \item $\mathbf{z}_{t_1}, \mathbf{z}_{t_2}, \ldots, \mathbf{z}_{t_{V+2}}$ are embeddings or feature representations of different views of the same data.
\item $L(\mathbf{z}_{t_1}, \mathbf{z}_{t_2}, \ldots, \mathbf{z}_{t_{V+2}})$ is the overall loss calculated based on these embeddings.
\item The outer summation $\sum_{i \in {1, 2}}{}$ iterates over two indices, typically representing pairs of global views.
\item The inner summation $\sum_{v=1}^{V+2}{}$ iterates over all the embeddings within each local view.
\item $\mathbf{1}_{v \neq i}$ is an indicator function that equals 1 if $v$ is not equal to $i$ (i.e., it ensures that we only consider embeddings from different views for comparison).
\item $l(\mathbf{z}_{t_v}, \mathbf{z}_{t_i})$ represents a similarity measurement between the embeddings $\mathbf{z}_{t_v}$ and $\mathbf{z}_{t_i}$.
\end{itemize}

% \begin{minted}{python}
% import torch
% import math

% def get_params(img, scale, ratio):
%     _, height, width = get_dimensions(img)
%     area = height * width
%     bone_ratio = 0
%     log_ratio = torch.log(torch.tensor(ratio))

%     while bone_ratio < 0.1:
%         target_area = area * torch.empty(1).uniform_(scale[0], scale[1]).item()
%         aspect_ratio = torch.exp(torch.empty(1).uniform_(log_ratio[0], log_ratio[1])).item()
        
%         w = int(round(math.sqrt(target_area * aspect_ratio)))
%         h = int(round(math.sqrt(target_area / aspect_ratio)))

%         if 0 < w <= width and 0 < h <= height:
%             i = torch.randint(0, height - h + 1, size=(1,)).item()
%             j = torch.randint(0, width - w + 1, size=(1,)).item()
%             temp_img = crop(img, i, j, h, w)
%             _, temp_h, temp_w = get_dimensions(temp_img)
%             total_pixels = temp_h * temp_w
%             one_layer = pil_to_tensor(temp_img)[0, :, :]
%             bone_pixels = torch.sum(torch.where(one_layer > 0, 1.0, 0.0)).item()
%             bone_ratio = bone_pixels / total_pixels
    
%     return i, j, h, w

% # Helper functions (not shown in pseudocode)
% # Define get_dimensions, crop, pil_to_tensor functions here

% \end{minted}
\begin{figure*}[ht]
      \centering
         \includegraphics[width=\textwidth]{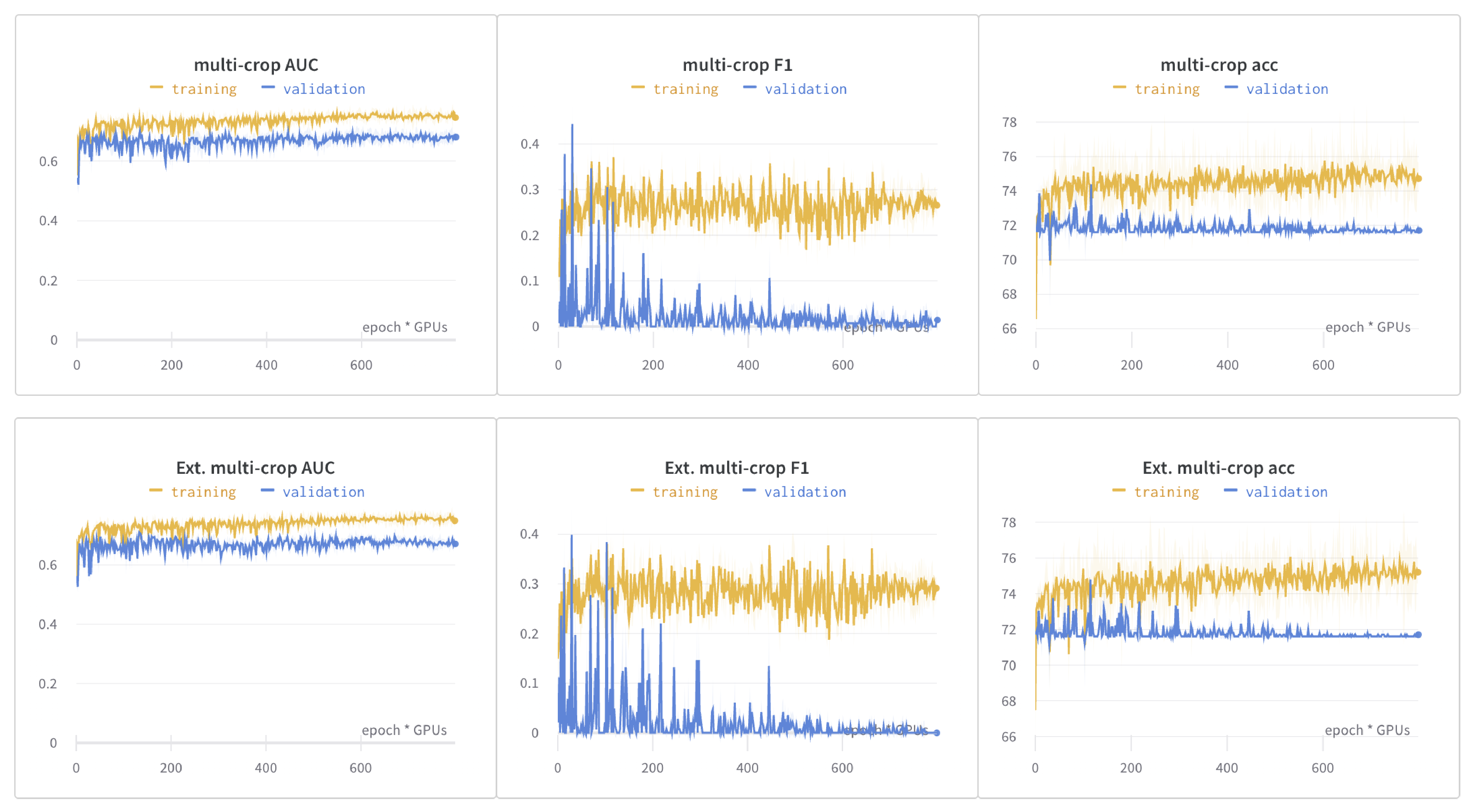}
         \caption{Metric score curves for supervised setting on training and validation data. }
         \label{fig:supervised}
\end{figure*}

\begin{figure*}[h]
      \centering
         \includegraphics[width=\textwidth]{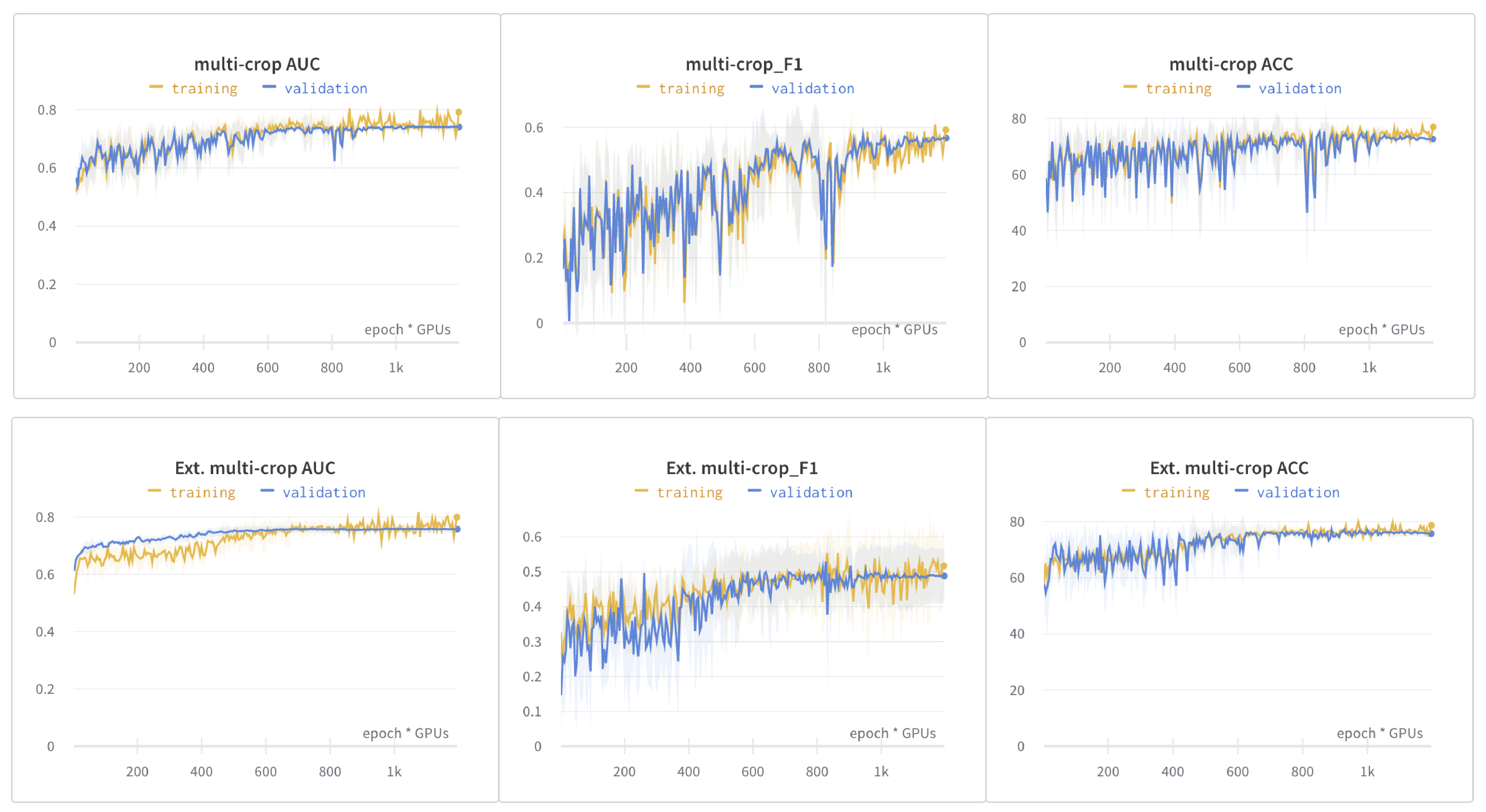}
         \caption{Metric score curves for SimCLR \citep{chen2020simple} on training and validation data.}
         \label{fig:SimCLR}
\end{figure*}

\begin{figure*}[h]
      \centering
         \includegraphics[width=\textwidth]{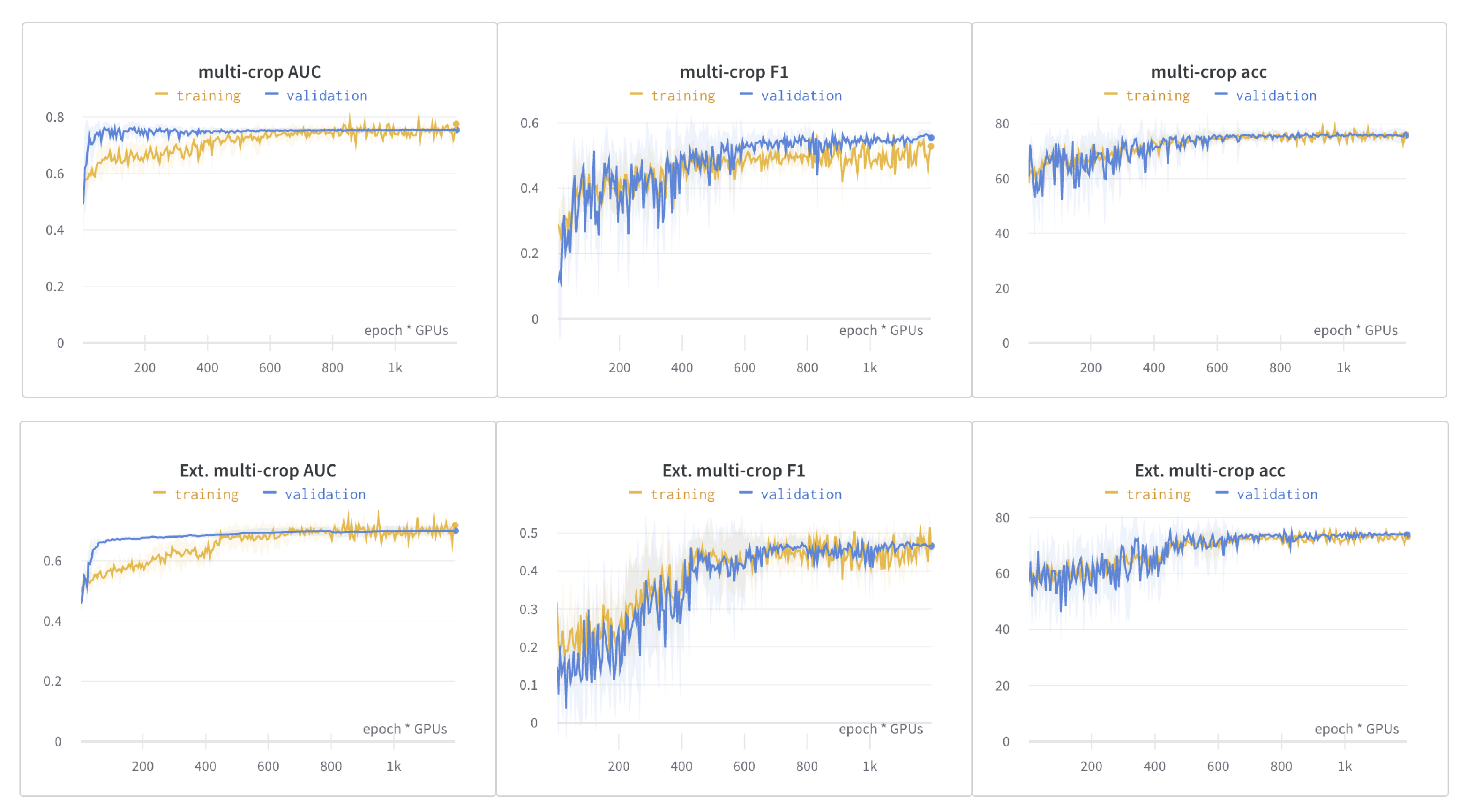}
         \caption{Metric score curves for SupCon \citep{khosla2020supervised} on training and validation data.}
         \label{fig:SupCon}
\end{figure*}

\begin{figure*}[h]
      \centering
         \includegraphics[width=\textwidth]{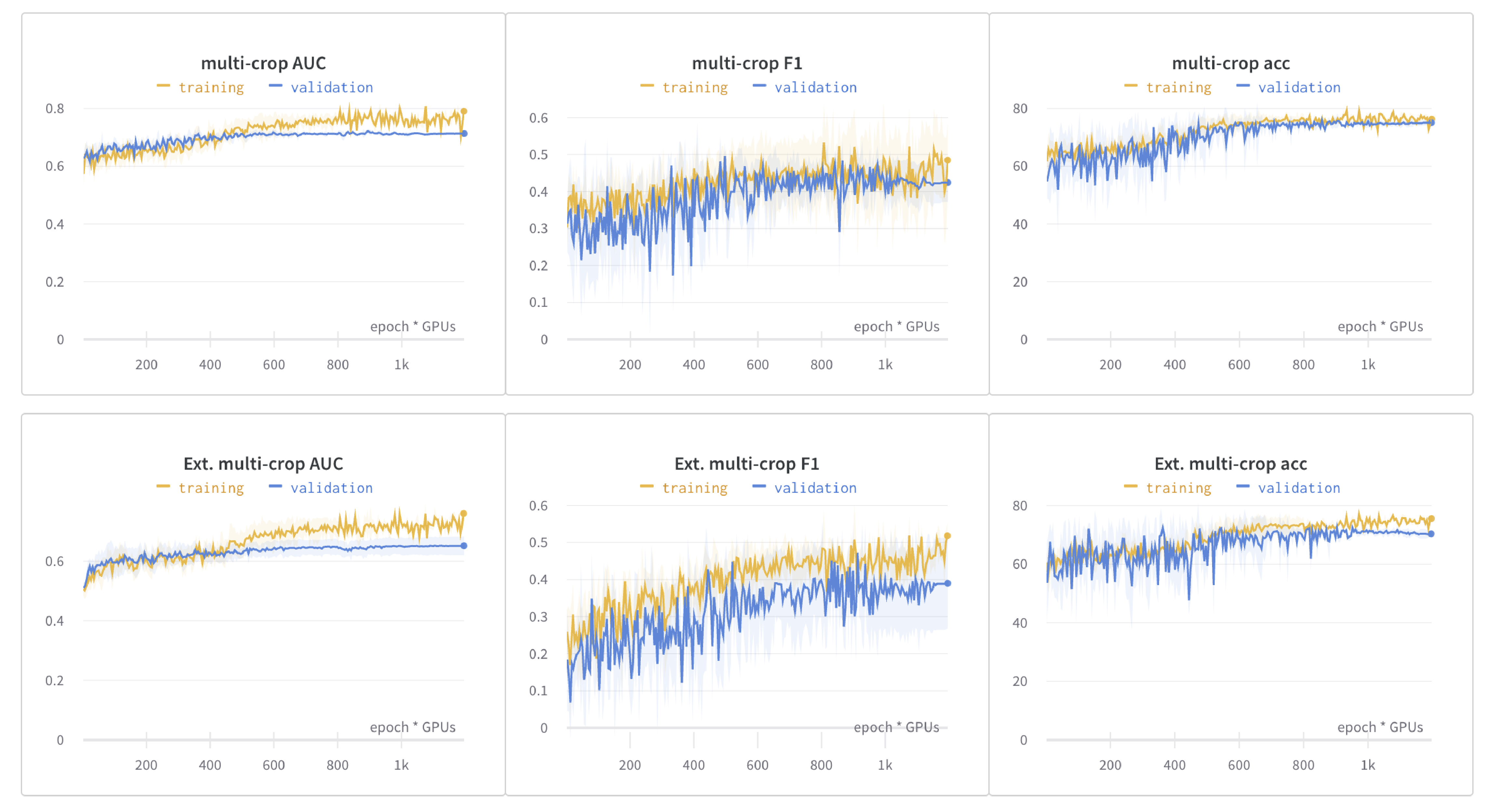}
         \caption{Metric score curves for SwAV \citep{caron2020unsupervised} on training and validation data.}
         \label{fig:SwAV}
\end{figure*}

\begin{figure*}[h]
      \centering
         \includegraphics[width=\textwidth]{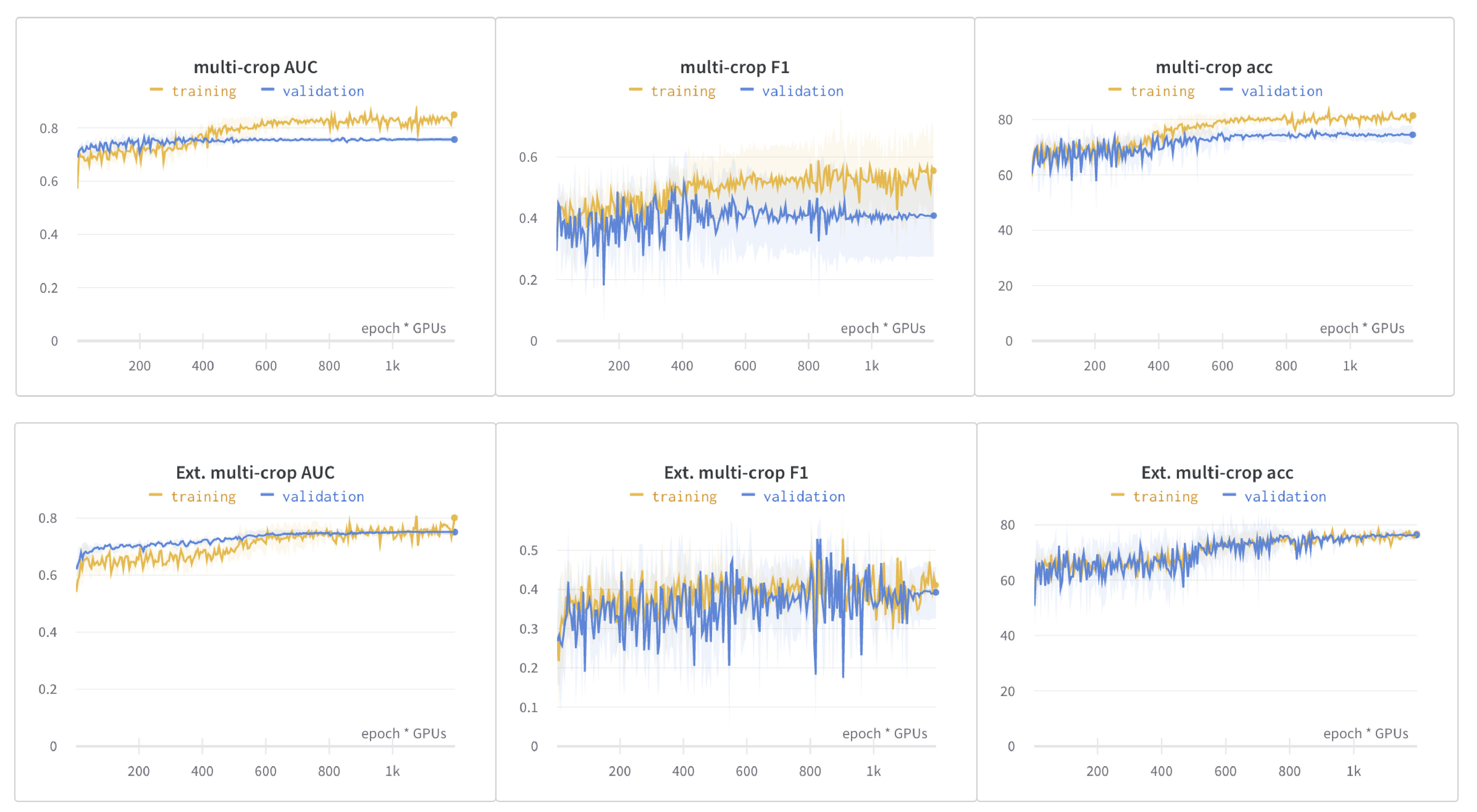}
         \caption{Metric score curves for VICReg \citep{bardes2021vicreg} on training and validation data.}
         \label{fig:VICReg}
\end{figure*}

\end{document}